\documentclass[graybox]{svmult}

\usepackage{graphicx}
\usepackage{float}
\usepackage{longtable}
\usepackage{tabularx}
\usepackage{tikz}
\usepackage{subfigure}
\usepackage[hidelinks]{hyperref}
\usepackage{enumitem}
\usepackage{makecell}
\newcommand{\RNum}[1]{\uppercase\expandafter{\romannumeral #1\relax}}

\usepackage[english]{babel}

\usepackage{mathtools}
\usepackage{xfrac}
\usepackage{multirow}
\usepackage{tikz}
\usepackage{wrapfig}
\usepackage{csquotes}
\usepackage{todonotes}
\usepackage{adjustbox}
\usepackage[noend,linesnumbered]{algorithm2e}
\usepackage{placeins}
\usepackage{array}
\usepackage{listings}
\usepackage[gen]{eurosym}
\usepackage{amsfonts}
\usepackage{xcolor}
\usepackage{enumitem}
\usetikzlibrary{shapes.misc}
\usetikzlibrary{positioning,petri,arrows.meta,shapes.geometric}
\DeclareRobustCommand{\circled}[1]{\tikz[baseline=(char.base)]{\node[shape=circle,draw,inner sep=0pt,minimum size=4.5mm] (char) {#1};}}

\definecolor{myred}{HTML}{e01b24}
\definecolor{mypurpleborder}{HTML}{c061cb}
\definecolor{mypurplefill}{HTML}{dc8add}
\definecolor{mybluefill}{HTML}{99c1f1}
\definecolor{myblueborder}{HTML}{3584e4}
\definecolor{mygrey}{HTML}{9a9996}
\definecolor{myorange}{HTML}{ff7800}

\DeclareRobustCommand{\isquare}[1]{\tikz[baseline=(char.base)]{\node[shape=rectangle,draw=myblueborder,fill=mybluefill,line width=0.7mm,inner sep=0pt,minimum size=3.1mm,font=\bf] (char) {\scriptsize #1};}}
\DeclareRobustCommand{\psquare}[1]{\tikz[baseline=(char.base)]{\node[shape=rectangle,draw=mypurpleborder,fill=mypurplefill,line width=0.7mm,inner sep=0pt,minimum size=3.1mm,font=\bf] (char) {\scriptsize #1};}}
\DeclareRobustCommand{\istar}[1]{\tikz[baseline=(char.base)]{\node[shape=star,draw=myorange,fill=myorange,line width=0.7mm,inner sep=0pt,minimum size=3.1mm,font=\bf] (char) {\scriptsize #1};}}

\newdimen\satlevel
\newdimen\satdiameter
\newcommand{\satisfaction}[2][]{%
    \satdiameter=2.5ex\relax
    \ifcase#2\relax
        \satlevel=0pt\relax
    \or
        \satlevel=0.125\satdiameter
    \or
        \satlevel=0.25\satdiameter
    \or
        \satlevel=0.375\satdiameter
    \or
        \satlevel=0.5\satdiameter
    \fi
    \tikz[baseline=-0.3\satdiameter]{%
        \draw[#1] (0,0) circle (0.5\satdiameter);
        \fill[#1] (0,0) circle (\satlevel);
    }%
}

\usepackage{graphicx}

\usepackage{comment}

\usepackage{csvsimple}

\newcommand{\orcidLM}	{\href{https://orcid.org/0000-0002-6866-0799}{\protect\includegraphics[scale=0.045]{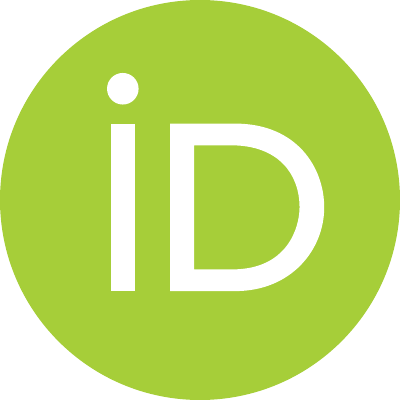}}}

\newcommand{\orcidRS}	{\href{https://orcid.org/0000-0003-1675-2592}{\protect\includegraphics[scale=0.045]{orcid.pdf}}}

\newcommand{\orcidRB}	{\href{https://orcid.org/0000-0002-5515-7158}{\protect\includegraphics[scale=0.045]{orcid.pdf}}}

\newcommand{\orcidBW}	{\href{https://orcid.org/0000-0002-6004-4860}{\protect\includegraphics[scale=0.045]{orcid.pdf}}}

\newcommand{\orcidJM}	{\href{https://orcid.org/0000-0002-6332-5801}{\protect\includegraphics[scale=0.045]{orcid.pdf}}}

\newcommand{\orcidJB}	{\href{https://orcid.org/0000-0002-3979-400X}{\protect\includegraphics[scale=0.045]{orcid.pdf}}}

\newcommand{\orcidJG}	{\href{https://orcid.org/0000-0001-7538-1248}{\protect\includegraphics[scale=0.045]{orcid.pdf}}}

\newcommand{\orcidYK}	{\href{https://orcid.org/0000-0001-6536-0785}{\protect\includegraphics[scale=0.045]{orcid.pdf}}}

\newcommand{\orcidFG}	{\href{https://orcid.org/0000-0003-0861-275X}{\protect\includegraphics[scale=0.045]{orcid.pdf}}}

\newcommand{\orcidSR}	{\href{https://orcid.org/0000-0001-5656-6108}{\protect\includegraphics[scale=0.045]{orcid.pdf}}}

\newcommand{\orcidMR}	{\href{https://orcid.org/0000-0003-2536-4153}{\protect\includegraphics[scale=0.045]{orcid.pdf}}}

\newcommand{\orcidME}	{\href{https://orcid.org/0000-0002-7739-9123}{\protect\includegraphics[scale=0.045]{orcid.pdf}}}

\newcommand{\orcidMF}	{\href{https://orcid.org/0000-0001-7030-282X}{\protect\includegraphics[scale=0.045]{orcid.pdf}}}

\newcommand{\orcidYB}	{\href{https://orcid.org/0000-0002-6407-7221}{\protect\includegraphics[scale=0.045]{orcid.pdf}}}

\newcommand{\orcidES}	{\href{https://orcid.org/0000-0001-7579-910X}{\protect\includegraphics[scale=0.045]{orcid.pdf}}}

\usepackage{type1cm}                                    %
\usepackage{makeidx}         \usepackage{graphicx}                                     \usepackage{multicol}        \usepackage[bottom]{footmisc}

\usepackage{newtxtext}       %
\usepackage[varvw]{newtxmath}       

\makeindex                                                           

\begin{document}

\title*{From Internet of Things Data to Business Processes: Challenges and a Framework}
\titlerunning{From Internet of Things Data to Business Processes}
\author{Jürgen Mangler\orcidJM,
Ronny Seiger\orcidRS,
Janik-Vasily Benzin\orcidJB, \\
Joscha Grüger\orcidJG,
Yusuf Kirikkayis\orcidYK,
Florian Gallik\orcidFG, \\
Lukas Malburg\orcidLM,
Matthias Ehrendorfer\orcidME,
Yannis Bertrand\orcidYB, \\
Marco Franceschetti\orcidMF,
Barbara Weber\orcidBW,
Stefanie Rinderle-Ma\orcidSR, \\
Ralph Bergmann\orcidRB,
 Estefanía Serral Asensio\orcidES, and
Manfred Reichert\orcidMR
}

\institute{Jürgen Mangler, Janik-Vasily Benzin, Matthias Ehrendorfer, Stefanie Rinderle-Ma \at Department of Informatics, Technical University of Munich, Garching, Germany,
\\\email{{juergen.mangler,janik.benzin,matthias.ehrendorfer,stefanie.rinderle-ma}@tum.de} \and
Ronny Seiger, Marco Franceschetti, Barbara Weber \at Institute of Computer Science, University of St.Gallen, St.Gallen, Switzerland
\\\email{{ronny.seiger,marco.franceschetti,barbara.weber}@unisg.ch} \and
Joscha Grüger, Lukas Malburg, Ralph Bergmann \at Business Information Systems II, University of Trier, Trier, Germany\\\email{{grueger,malburgl,bergmann}@uni-trier.de} \and
Joscha Grüger, Lukas Malburg, Ralph Bergmann \at German Research Center for Artificial Intelligence (DFKI), Branch University of Trier, Germany \\ \email{{joscha.grueger,lukas.malburg,ralph.bergmann}@dfki.de} \and
Yusuf Kirikkayis, Florian Gallik, Manfred Reichert \at Institute of Databases and Information Systems, Ulm University, Ulm, Germany
\\\email{{yusuf.kirikkayis,florian-1.gallik, manfred.reichert}@uni-ulm.de} \and
Yannis Bertrand, Estefanía Serral Asensio \at Research Centre for Information Systems Engineering (LIRIS), KU Leuven, Leuven, Belgium,
\\\email{{yannis.bertrand, estefania.serralasensio}@kuleuven.be}
}

\authorrunning{J. Mangler et al.}

\maketitle

\abstract{The IoT and Business Process Management (BPM) communities co-exist in many shared application domains, such as manufacturing and healthcare. The IoT community has a strong focus on hardware, connectivity and data; the BPM community focuses mainly on finding, controlling, and enhancing the structured interactions among the IoT devices in processes. While the field of Process Mining deals with the extraction of process models and process analytics from process event logs, the data produced by IoT sensors often is at a lower granularity than these process-level events. The fundamental questions about extracting and abstracting process-related data from streams of IoT sensor values are: (1)~Which sensor values can be clustered together as part of process events?, (2)~Which sensor values signify the start and end of such events?, (3)~Which sensor values are related but not essential?
This work proposes a framework to semi-automatically perform a set of structured steps to convert low-level IoT sensor data into higher-level process events that are suitable for process mining. The framework is meant to provide a generic sequence of abstract steps to guide the event extraction, abstraction, and correlation, with variation points for plugging in specific analysis techniques and algorithms for each step. To assess the completeness of the framework, we present a set of challenges, how they can be tackled through the framework, and an example on how to instantiate the framework in a real-world demonstration from the field of smart manufacturing. Based on this framework, future research can be conducted in a structured manner through refining and improving individual steps.}


\section{Introduction}
\label{sec:intro}

In IoT environments, large amounts of procedural data are generated from IoT devices, information systems, and other software applications. The use of this data can foster the development of innovative applications in process control~\cite{messner2019closed,SEIGER2022575,10.1007/978-3-030-66498-5_8,in4pl20,Hoffmann.2022_ProGAN,Malburg_AdaptiveWorkflowManagement_2022,kirikkayis2023integrating,pauker2018centurio}, process conformance checking~\cite{ehrendorfer2019conformance,stertz2020data,stertz2020temporal,franceschetti2023proambition}, and process enhancement~\cite{pauker2021industry,cs6099}, among others. Particularly, the use of process mining techniques to analyze not only process data but also IoT-collected data could provide important insights into processes and interactions as shown in different applications in the manufacturing domain, such as~\cite{mangler2014cpee,SEIGER2022575,10.1007/978-3-030-66498-5_8,cs6099,pauker2021industry}. In these applications, IoT actuators are used to realize and execute process activities, while IoT sensors and smart tags are used to closely monitor the execution environment and involved resources~\cite{Sisinni.2018_IndustrialIoT,SEIGER2022575,Elsaleh.2019_IoTStream,janiesch2020internet,in4pl20}. IoT technology can therefore capture the context in which certain process tasks are performed, allowing process mining techniques to better understand and analyze the processes~\cite{10.1007/978-3-030-98581-3_8,seiger2020towards,brzychczy2024process}.

As such, besides the procedural data generated from the process execution systems, the data captured by IoT should also be considered an integral part of the process execution in the form of \emph{IoT-enriched event logs}~\cite{fi15030109,Malburg.2022_IoTEnrichedEventLog}. Both the procedural nature of sensor logs, and the tight integration of these with the process executions and the executing resources~\cite{ehrendorfer2021sensor} makes sensor data an integral part of process-based application scenarios in IoT~\cite{seiger2020towards,SEIGER2022575,10.1007/978-3-030-98581-3_8}.

However, the integration of IoT data and process data to be used for process mining is still often done ex-post in a manual fashion during a separate pre-processing phase~\cite{weyers2022method,weyers2022,Malburg.2022_IoTEnrichedEventLog}. In these cases, the data from the IoT environment is still collected and stored separately, and only later it is explicitly connected to the notion of a process, which is non-trivial as pointed out in the challenge ``Bridging the Gap Between Event-based and Process-based Systems'' in the BPM-IoT manifesto~\cite{janiesch2020internet}. In case no explicit process orchestration or monitoring happens via a BPM system, and thus no explicit process data is available~\cite{franceschetti2023event}, the IoT data is the sole provider of process-related data, which has to be separately identified, extracted, abstracted, and correlated~\cite{diba2020extraction,weyers2022,seiger2020towards}.

We define the following two extreme states of a spectrum that IoT data sets may be in regarding their inclusion of process-related data (called \textbf{Process Awareness}), with certain levels of process awareness in between:

\begin{itemize}

  \item An IoT data set (log) exhibits \textbf{Full Process Awareness}, when all relations between sensor events and process events are known.

  \item An IoT data set (log) exhibits \textbf{No Process Awareness}, when it only contains raw sensor events, and no process events at all.

\end{itemize}

In between \emph{full process awareness} and \emph{no process awareness}, IoT data sets may exhibit some known correlations with the process executions, e.g., the type of a process activity that a specific sensor event belongs to might be known, but not the specific instance, or the correlation of a process-level event identified from the sensor events with a specific process instance might be unknown~\cite{weyers2022method}.

In this work, we propose a framework to transition IoT data sets from \textbf{No Process Awareness} to \textbf{Full Process Awareness}, in a structured, non-exploratory fashion.
The contribution of the paper is three-fold:

\begin{itemize}

    \item We highlight relevant challenges regarding the automatic derivation and connection of process data with IoT data to allow for a better subsumption of future work in this area of research.

    \item We propose a generic framework to address the identified challenges and to derive process data from IoT data step-by-step to facilitate process mining.

    \item We showcase the framework using a real-life example from smart manufacturing.

\end{itemize}

This work is structured as follows: Section~\ref{sec:awareness} introduces the notion of \emph{Process Awareness} in detail. Section~\ref{sec:rel} discusses related work. Section~\ref{sec:challenges} presents challenges regarding the connection of process and IoT data. Section~\ref{sec:contribution} presents our framework to (semi-)automatically increase the level of process awareness in IoT data sets with a real-world example from smart manufacturing. Section~\ref{sec:conclusion} concludes the paper and shows opportunities and starting points for future work.

\section{IoT Data and Process Awareness}
\label{sec:awareness}

\begin{figure}
  \centering
  \includegraphics[width=0.8\linewidth]{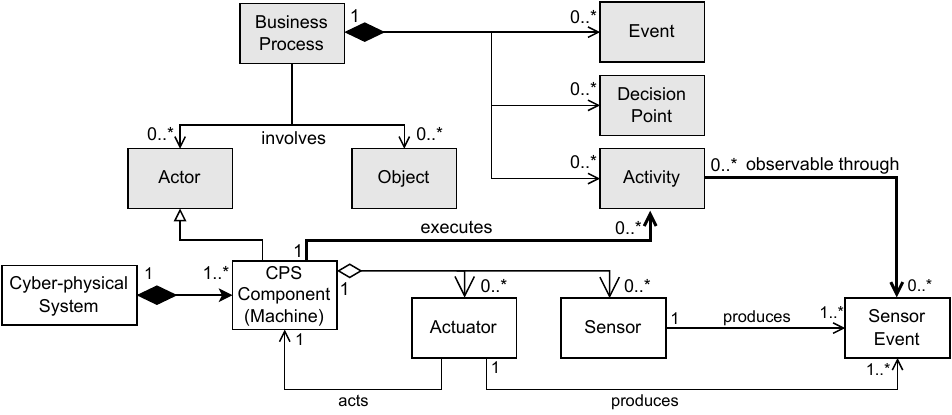}
  \caption{Relation between concepts from BPM (in grey) and CPS/IoT (adopted from \cite{weyers2022}).}
  \label{fig:activitymachinesensor}
\end{figure}

This work aims at bringing the world of BPM together with the world of IoT and cyber-physical systems (CPS). The conceptual model linking the worlds of BPM (highlighted in grey) with IoT/CPS that we base our work on is depicted in Fig.~\ref{fig:activitymachinesensor}. A \emph{Business Process} contains \emph{Events}, \emph{Decision Points} and \emph{Activities}~\cite{dumas2013fundamentals}. It involves \emph{Actors} playing an active part in the process and \emph{Objects}. In the context of our work, components of a CPS (e.g.,~a production machine) can also be actors that are able to execute process activities, which in turn can be observed through \emph{Sensor Events}.

\begin{figure}
  \centering
  \includegraphics[width=0.75\linewidth]{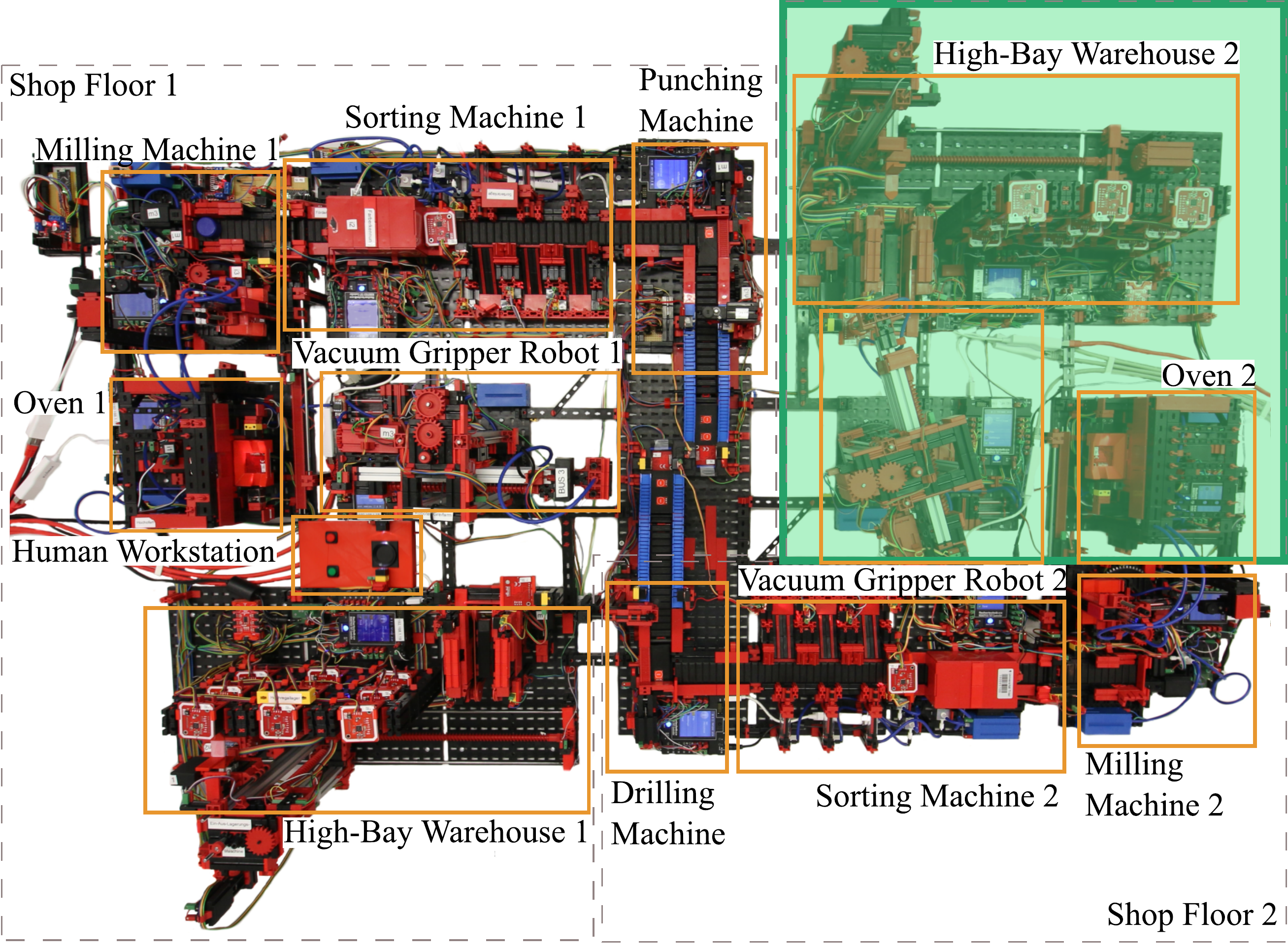}
  \caption{Factory model representing a CPS with highlighted CPS Components~\cite{10.1007/978-3-030-66498-5_8}.}
  \label{fig:factory}
\end{figure}

IoT data is typically generated by a set of sensors that measure a set of values emitted as \emph{Sensor Events}~\cite{bauer2013iot}. These sensors can be standalone or part of a larger CPS component (e.g., a production machine or production cell) resulting in a particular \emph{sensor topology} for a CPS. Actuators interact with the physical world and manipulate its state. As actuators may also produce data regarding their current state and actions, we consider data produced from actuators as relevant \emph{Sensor Events} for our analysis, too. Fig.~\ref{fig:factory} shows a Fischertechnik model of a smart factory  (i.e., an implementation for simulation purposes) that we use for research\footnote{\url{https://www.fischertechnik.de/en/products/industry-and-universities/training-models}, last access: 2024-05-13}. It represents a CPS consisting of several (highlighted) CPS components that again consist of a multitude of sensors and actuators. Sensor events may also be directly related with actors and objects. However, we put process-related concepts in the center of our investigations, which is why the key relationship for our work to bridge the BPM and IoT worlds is the \emph{observable through} association between \emph{Activity} and \emph{Sensor Event} in Fig.~\ref{fig:activitymachinesensor}.

The sensor events are typically either directly pushed to some IoT data log by the sensor, or collected and stored in an IoT data log by an external mechanism. Thus, a set of IoT data logs, given some data retention policies, holds historical information regarding all timestamped observations from a set of sensors. For each sensor, the stream of events (called \emph{Raw Sensor Stream} in this work) is available as a time series~\cite{8926446}.

\subsection{Process Activity--Sensors--Machines Relationship}

When associating data collected from IoT with process data to achieve full process awareness, one has to consider that sensors can be related to process activities without explicitly knowing the corresponding actions, i.e., which machine or human triggered a certain chain of events. This is because one activity might trigger, e.g., the start of a machine, while other activities might fine-tune the parametrization of the machine at runtime. A third activity might then wait for an event signifying the end of the operation. This also means that one sensor at any given time might produce data relevant to many activities, while one activity might be enacted in the context of many sensors (cf.~Fig.~\ref{fig:activitymachinesensor}). The same holds true for machines: a sensor might at any given time produce data relevant to the operation of many machines (e.g., a temperature sensor might deliver data relevant to the operation of multiple machines), while one machine might operate in the context of multiple sensors~\cite{cs6099,weyers2022,9860216,klein2019ftonto}.

The reason for collecting IoT data typically is to monitor the effects of a series of actuations, e.g., as part of a production process, the treatment of a patient, or as part of a building automation scenario. The collected data can be used to~\cite{10.1007/978-3-030-58666-912,seiger2020towards,9860104,janiesch2020internet,in4pl20,fi15030113}:

\begin{itemize}

    \item monitor, visualize, and track progress,
    \item find deviations from rules,
    \item find drifts in relation to goal functions.

\end{itemize}

While these analysis targets are very generic and universally applicable, they can only be reached with detailed knowledge of the scenario, its rules, goal functions, and the underlying \textbf{Business Process}~\cite{dumas2013fundamentals}.

All IoT scenarios have in common that a series of actuations, potentially influenced by humans and the environment, are based on a blueprint that defines when something should happen, and which order relations might exist between actuations. This blueprint, the \textbf{Process}~\cite{janiesch2020internet,dumas2013fundamentals}, might either be:

\begin{itemize}

    \item Implicit, i.e., the result of the interactions of some hard-coded software components with potential human involvement, or

    \item Explicit, i.e., a known artifact (e.g., the \textbf{Process Model}), which is the basis for enacting a certain scenario, e.g., through a process engine.

\end{itemize}

The relationship between IoT data and processes can be understood as IoT data tracking changes in the observed environment while processes and activities are executed, and thus allow to observe their effects~(cf.~Fig.~\ref{fig:activitymachinesensor}~\cite{bauer2013iot}).

Albeit, some caveats to complicate the situation exist:

\begin{itemize}

    \item Sensors, e.g., temperature sensors, might collect data while no scenario is enacted. Thus, data might not be relevant for the process~\cite{weyers2022}.

    \item Not all sensor data can be directly assigned to a specific scenario, only the combination of a set of individual observations from different IoT data logs (e.g., temperature sensors combined with the status of an oven) might allow identifying significance for a particular scenario~\cite{seiger2020towards}.

    \item As IoT data from different sources is available in different granularity, substantial effort for aggregation of the data has to be made~\cite{soffer2019event,seiger2020towards}.

\end{itemize}

\begin{figure}
  \centering
  \includegraphics[width=0.8\linewidth]{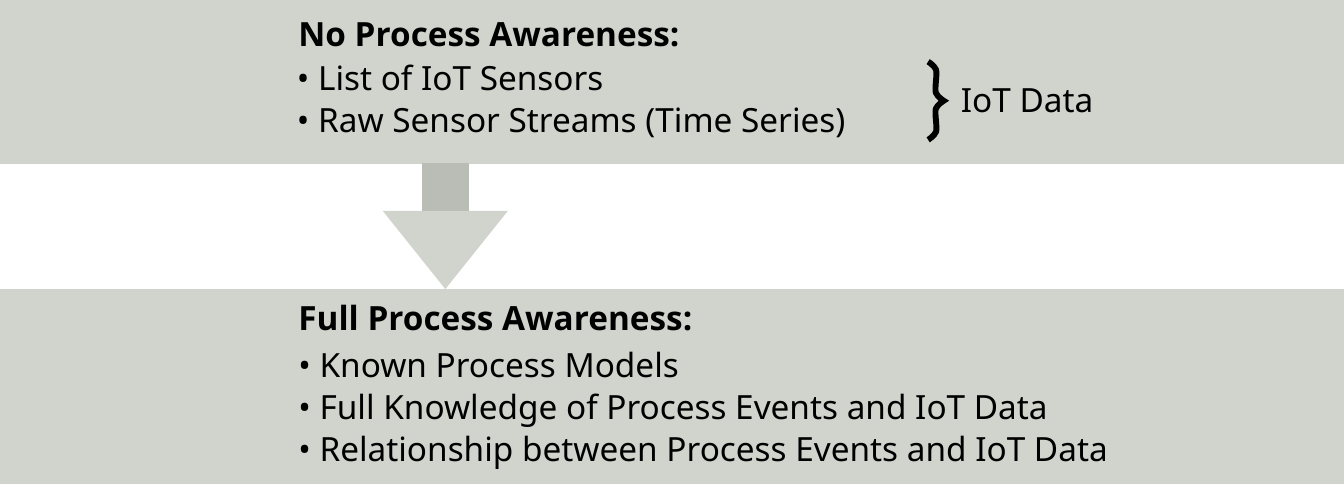}
  \caption{Relation between IoT data and process awareness.}
  \label{fig:layers-simple}
\end{figure}

Consequently, as briefly introduced in Section~\ref{sec:intro}, we define \textbf{No Process Awareness} as a situation where only raw sensor streams from a set of data sources are available (cf.~Fig.~\ref{fig:layers-simple})~\cite{weyers2022}. \textbf{Full Process Awareness}, on the other hand, \textbf{is the goal} as it represents the full contextualization of all data regarding a certain scenario, which is represented explicitly by a set of process models~\cite{cs6099}. Each event in each available sensor stream time series has been either discarded due to irrelevance or assigned to an observation of an actuation. Each actuation can be related to many process-level events, and these events might be related to multiple process activities~\cite{https://doi.org/10.48550/arxiv.2206.11392,weyers2022}. For the remainder of this section, we further explore the two extreme states of the Process Awareness spectrum with different levels in between.

\subsection{Example: Smart Manufacturing Process}
\label{sec:example}

\begin{figure}
  \centering
  \includegraphics[width=1\linewidth]{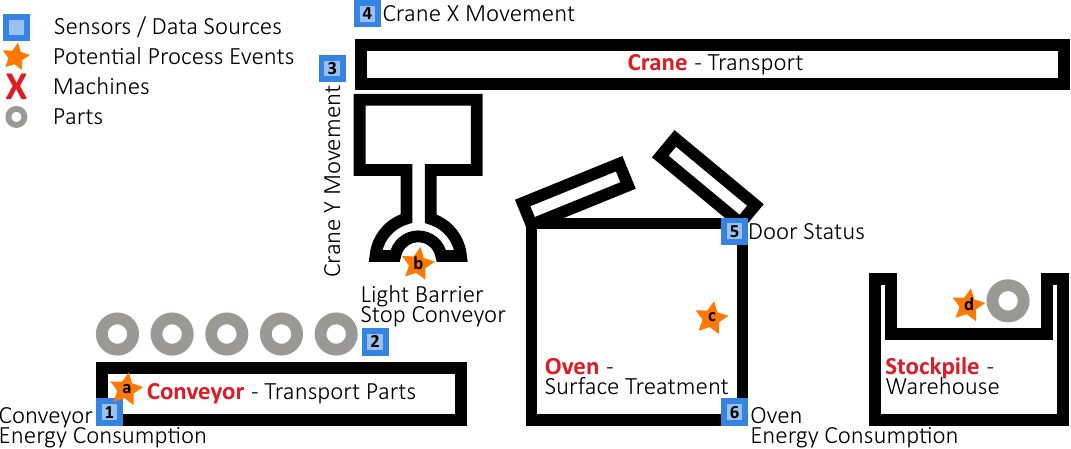}
  \caption{Running example, part of factory simulation model shown in Fig.~\ref{fig:factory}.}
  \label{fig:running}
\end{figure}

In order to better exemplify the concepts discussed in this section, and following sections, for the remainder of this work we will focus on a specific slice of the scenario displayed in \textbf{Fig. \ref{fig:factory} (top right area, green box)}. This \textbf{running example} is displayed in simplified form in \textbf{Fig. \ref{fig:running}} and describes the interaction between a \textcolor{myred}{Conveyor} which transports raw \textcolor{mygrey}{parts} to a \textcolor{myred}{Crane} (``Vacuum Gripper Robot 2''), which can move the parts to an \textcolor{myred}{Oven} (``Oven 2'') for surface treatment, and subsequently to a \textcolor{myred}{Stockpile} (``High-Bay Warehouse 2'').

Our running example utilizes a total of 6 different sensors, respectively data sources, which deliver the following values:

\begin{enumerate}[leftmargin=30pt]

    \item[\isquare{1}] is a continuous sensor that emits the energy consumption levels of the \textcolor{myred}{Conveyor} in watts every second. When the \textcolor{myred}{Conveyor} is not running, the sensor emits the value (close to) $0$ every second.

    \item[\isquare{2}] is a binary sensor that switches from $0$ to $1$ whenever a \textcolor{mygrey}{part} is in its field. It is also used by the internal programming of the factory to start/stop the \textcolor{myred}{Conveyor} to ensure that whenever a \textcolor{mygrey}{part} is taken by the \textcolor{myred}{Crane}, a new part is moved into a position where it can
    be picked up.

    \item[\isquare{3}] is a continuous sensor that monitors the y position of the \textcolor{myred}{Crane}, i.e., how far the gripper is extended downwards. It delivers values
    every second whenever the position changes, it delivers no values, when the position is unchanged. Fully extended the y value for the \textcolor{myred}{Crane} is $100$, fully retracted the value is $0$.

   \item[\isquare{4}] is a continuous sensor that monitors the x position of the \textcolor{myred}{Crane}, i.e., in which position above the \textcolor{myred}{Conveyor}, \textcolor{myred}{Oven}, or \textcolor{myred}{Stockpile} the crane is. It delivers values every second whenever the position changes, it delivers no values, when the position is unchanged. When the \textcolor{myred}{Crane} is above the \textcolor{myred}{Conveyor} the value is $0$, when it is above the \textcolor{myred}{Oven} it is $30$, and when it is above the \textcolor{myred}{Stockpile} it is $60$.

   \item[\isquare{5}] is a binary sensor that monitors if the door of the \textcolor{myred}{Oven} is opened ($1$) or closed ($0$).

   \item[\isquare{6}] is a continuous sensor that emits the energy consumption levels of the \textcolor{myred}{Oven} in watts every second. When the \textcolor{myred}{Oven} is not running, the sensor emits the value (close to) $0$ every second. As the \textcolor{myred}{Oven} is always kept at temperature, positive energy consumption does not necessarily mean that a part is currently
   processed.

\end{enumerate}

Between all of these sensor values, a wide range of process events could be specified, leading to more or less fine-granular process models that describe the interaction between the \textcolor{myred}{Conveyor}, the \textcolor{myred}{Crane}, the \textcolor{myred}{Oven}, and the \textcolor{myred}{Stockpile}. Potential process events that we want to highlight for our
running example are:

\begin{enumerate}[leftmargin=30pt]

    \item[\istar{a}] The \textcolor{myred}{Conveyor} moves a \textcolor{mygrey}{part} into position, which can be deducted from \isquare{1}.

    \item[\istar{b}] The \textcolor{myred}{Crane} grabs a \textcolor{mygrey}{part}, which can be deducted from certain values of \isquare{2}, \isquare{3}, and \isquare{4}.

    \item[\istar{c}] A \textcolor{mygrey}{part} is surface treated in the \textcolor{myred}{Oven}. If \isquare{3}, \isquare{4} have certain values, and \isquare{5} is 1, then the oven is currently being loaded. If \isquare{3}, \isquare{4} have certain values, and \isquare{5} is 0, then the oven is processing a \textcolor{mygrey}{part}.

    \item[\istar{d}] The \textcolor{myred}{Crane} drops a \textcolor{mygrey}{part} into the \textcolor{myred}{Stockpile}, which can be deducted from certain values of \isquare{3}, and \isquare{4}.

\end{enumerate}

\subsection{No Process Awareness}
\label{subsec:noprocessawareness}

This state of Process Awareness typically occurs in brownfield scenarios. IoT data collected from existing complex IoT environments is partially analyzed regarding certain domain-specific preordained, non-process related questions. Whenever new questions arise, the data aggregation is refined to derive the required answers. This state is typically prone to breakages as the following properties might hold true:

\begin{itemize}

  \item Unknown Side-Effects: During the operation of the IoT environment, actuations might occur which are not foreseen in the data analysis. The software steering the actuations, or the human's or environment's influence on the process might not be fully understood~\cite{seiger2019toward,marrella2018supporting}.

  \item Breaking Changes: Such side-effects might come into existence whenever changes to the IoT environment are made. Changes not only include the addition of new actuators or sensors, but also the (software) upgrade of existing ones~\cite{6983801}.

\end{itemize}

Consequentially, all changes have to be tracked, monitored, and all data analysis tasks have to be rechecked if they still deliver the required answers, or have to be adapted to do so. We can sum up the properties of this state as follows, while not all of them must apply to all scenarios~\cite{seiger2020towards,soffer2019event,SEIGER2022575,monostori2014cyber,weyers2022,franceschetti2023characterisation,zerbato2021granularity,zerbato2021granularityws}:

\begin{itemize}

    \item Hard-coded software.
    \item No or only partial knowledge of the process.
    \item Event correlation ambiguities: no clear assignment of events to individual actuations.
    \item Partial monitoring and tracking of data.
    \item Partial knowledge about the dependencies between sensors (i.e., sensor topology), and how they influence different scenarios.
    \item Incomplete data, i.e., sensors might be missing to monitor relevant aspects of the process executions.
    \item Inconsistent data resulting from different data granularities of different sensors.

\end{itemize}

Table~\ref{tab:logno} shows an excerpt from a typical IoT data log referring to the example scenario described in Section~\ref{sec:example} which only contains sensor readings with a minimal set of sensor attributes, sensor value, and timestamp of the sample. There is no process-related information available.

\begin{table}
    \centering
    \caption{Excerpt from an IoT data log with no process awareness.}
    \begin{tabular}{|c|c|c|c|c|c|c|c|c|}
        \hline
        & \makecell{Sensor\\ID} & \makecell{Sensor\\value} & \makecell{Sensor\\timestamp} & \makecell{Case} & \makecell{Process\\event ID} & \makecell{Event\\timestamp} & \makecell{Lifecycle\\transition} & \makecell{Label} \\ \hline
        1 & \makecell{\isquare{1}} & 512 & 1 & -- & -- & -- & -- & --\\ \hline
        2 & \makecell{\isquare{2}} & 0   & 1 & -- & -- & -- & -- & --\\ \hline
        3 & \makecell{\isquare{1}} & 0   & 2 & -- & -- & -- & -- & --\\ \hline
        4 & \makecell{\isquare{2}} & 1   & 2 & -- & -- & -- & -- & --\\ \hline
        5 &\makecell{\isquare{3}} & 100 & 3 & -- & -- & -- & -- & --\\ \hline
        6 &\makecell{\isquare{4}} & 0   & 3 & -- & -- & -- & -- & --\\ \hline
        7 & ... & ... & 4 & -- & -- & -- & -- & --\\ \hline
    \end{tabular}

    \label{tab:logno}
\end{table}

\subsection{Full Process Awareness}
\label{subsec:fullprocessawareness}

This state of Process Awareness typically occurs in greenfield scenarios, when new IoT environments are planned and realized. Furthermore, it might occur in brownfield scenarios, when process management software is introduced as part of an automation effort, such as~\cite{cs6099,8432166}. Full process awareness is often characterized by~\cite{cs6099,https://doi.org/10.48550/arxiv.2206.11392,SEIGER2022575,10.1007/978-3-030-66498-5_8,Malburg.2022_IoTEnrichedEventLog,fi15030113}:

\begin{itemize}

    \item Full knowledge of the relevant processes in the form of a set of process models.

    \item Imperative control or at least full monitoring through a process engine.

    \item Sensors which are linked to process activities, i.e., sensors might contribute data to one or more activities.

    \item Complete Data: if data is missing, the process is not specified well enough.

    \item Logs, which are automatically generated as part of the execution, and containing events and process data at a suitable, homogeneous level of granularity~\cite{zerbato2021granularityws}; complemented by IoT data~\cite{fi15030109}.

\end{itemize}

\begin{table}
    \centering
    \caption{Excerpt from an IoT-enriched event log with full process awareness related to control flow.}
    \begin{tabular}{|c|c|c|c|c|c|c|c|c|}
        \hline
        & \makecell{Sensor\\ID} & \makecell{Sensor\\value} & \makecell{Sensor\\timestamp} & \makecell{Case} & \makecell{Process\\event ID} & \makecell{Event\\timestamp} & \makecell{Lifecycle\\transition} & \makecell{Label} \\ \hline
        1 & \isquare{1} & 512 & 1 & part 1 & \istar{a} & 1 & start & move part \\ \hline
        2 & \isquare{1}, \isquare{2} & 0, 1 & 2, 2 & part 1 & \istar{a} & 2 & complete & move part \\ \hline
        3 & \isquare{2}, \isquare{3}, \isquare{4} & 1, 100, 0 & 2, 3, 3 & part 1 & \istar{b} & 3 & start & grab part \\ \hline

        4 & ... &  ... &  ... & part 1 & ... & ... & ... & ... \\ \hline
    \end{tabular}

    \label{tab:logfull}
\end{table}

In this case, data stores for individual sensors might still exist, but they are no longer used because all data is collected as part of the execution of process instances stored in an IoT-enriched process event log~\cite{https://doi.org/10.48550/arxiv.2206.11392}. Table~\ref{tab:logfull} contains the same sensor readings as Table~\ref{tab:logno}, but these are now fully contextualized in the control flow of the process execution (cf.~scenario process in Section~\ref{sec:example}) as indicated by the additional process attributes according to the XES standard~\cite{7740858}. In this example, the sensor events are clearly related to specific process events for one case, i.e., an executed process instance, namely the workpiece ``part 1''. Furthermore the following important concepts are exemplified in Table~\ref{tab:logfull}:

\begin{itemize}

  \item One or multiple sensors and thus one or multiple sensor events might contribute to one process event.

  \item Row 1 and row 2 describe different events in the lifecycle of the same process task (start/complete of \emph{move part}). Multiple combinations of sensor events might point to the same process task, i.e., in our example \isquare{1} is sufficient for start, but the combination of \isquare{1} and \isquare{2} with different values point at the and also is more robust (i.e., more sensors involved means better distinction).

\end{itemize}

Thus it becomes clear, that a process event is defined by a combination of sensors and the specific values they have at a given point in time.

~\newline\noindent{}But how are the sensor values collected? We differentiate between three options of realizing the data collection.

\subsubsection{Option 1: Explicit Modeling of Sensor Data Collection}

This way is about modeling the process in the most fine-grained and detailed way possible, including the explicit collection of sensor data as process tasks (activities) that are part of the process model~\cite{cs6099}. This has the advantage that

\begin{itemize}

    \item every process engine can support it,

    \item and models are better to follow as everything is made explicit in an already existing modeling language (e.g., BPMN~2.0).

\end{itemize}

On the downside, the resulting models are complex and complicated. Furthermore, the data collection might be hindered by the expressiveness of the modeling language, e.g., BPMN~2.0 requires complex synchronization modeling if data relevant to multiple in-parallel running tasks has to be collected in separate process instances~\cite{10.1007/978-3-030-58638-6_2}. While this downside can be alleviated by modeling the process in a more abstract way (fewer tasks, higher abstraction, ``more knowledge'' in lower levels), this requires increased effort to maintain a good modeling quality and decreases the level of detail of monitoring.

\subsubsection{Option 2: Implicit IoT Data Collection}

In this approach, the process model contains additional information/annotations that allow data collection implicitly, independent of the traditional process execution, but still automatically. For example, if process activities contain information about linked sensors, and how to collect data from them, this data collection can be enacted automatically in parallel to the normal process execution~\cite{in4pl20,9860216}.

The main advantage is that the process models stay simple, but on the downside some additional knowledge describing the connection to the IoT environment (e.g., in the form of annotations~\cite{in4pl20,9860216}) has to be provided.

\subsubsection{Option 3: Raw IoT Data Collection}

In this approach, the process notion is completely ignored, and data is collected directly from all sensors. This is the most common way of data collection. In contrast to the two other collection methods, where the data can be used for compliance checking and process refinement, here the goal is to discover and create a new process model. The downside is that for
getting the correct granularity of the process models, some of the steps in the framework proposed in this paper (cf.~Section~\ref{sec:contribution}) become much more reliant on domain expert input and validation~\cite{weyers2022}.

\section{Related Work}
\label{sec:rel}

Augmenting business processes with IoT technologies brings several benefits, but does not come without challenges~\cite{janiesch2020internet,10.1007/978-3-030-85469-0_1,de2024interplay}. In this section, we discuss related research addressing IoT data processing as well as event extraction and event abstraction~\cite{diba2020extraction} in the context of IoT with the goal of increasing the level of process awareness associated with IoT data, and ultimately, to enable process mining~\cite{akhramovich2024systematic}.

One of the main benefits is the improved recognition of activities and processes from sensor data, e.\,g.,~with the help of multi-modal approaches such as the one by Rebmann et al.~\cite{rebmann2019enabling} that combines motion and vision sensors with user feedback for detecting and disambiguating known activity types/tasks in real-time. Other approaches employ camera-based object detection to track workpieces within a smart factory and correlate them to the execution of individual process activities~\cite{Malburg.2021_ObjectDetection}. Activity recognition is often combined with process mining for extracting knowledge to analyze and optimize the underlying processes~\cite{mannhardt2018taxonomy,9079475}. Applying process mining, in turn, requires bridging the gap between low-level sensor data and the event logs needed for process mining--a well-known issue, which gives rise to succeeding challenges~\cite{janiesch2020internet}. 

The data used for process mining can reside in different data sources. \emph{Event Extraction} includes the identification of the event data and its extraction~\cite{diba2020extraction}. 
To leverage the opportunities IoT brings for improved recognition of (manual) activities and processes from sensor data~\cite{soffer2019event}, differences in granularity at which the data is recorded and at which it shall be analyzed~\cite{diba2020extraction} need to be bridged, also referred to as \emph{Event Abstraction}. In turn, \emph{Event Correlation} aims at associating event data extracted from data sources to \emph{cases} of a process~\cite{brzychczy2024process}. An overview of non-IoT related event abstraction techniques in the field of process mining is provided in a literature review by van Zelst~\cite{van2021event}. A paper by Diba et al.~\cite{diba2020extraction} provides an overview of different methods for event abstraction and correlation in the traditional process mining context. 

\subsection{Event Extraction}

Identifying relevant data for process mining and extracting it from different sources is part of \emph{Event Extraction}~\cite{diba2020extraction}.  Existing approaches use database schema, process documents, domain models, event models, and domain knowledge for the identification and extraction of relevant data (cf.~\cite{diba2020extraction,jans2019building}) and mostly focus on the extraction of event data from databases and traditional information systems. Recently, several works looked into the use of IoT as data source for process mining (e.g.,~\cite{akhramovich2024systematic,7549355,10.1007/978-3-030-58666-912,Koschmider2020framework,ehrendorfer2019conformance,10.1007/978-3-030-98581-3_8,li2023rectify}). Similarly, in our work we are moving towards IoT sensors as new data sources for event extraction and process correlation~\cite{seiger2020towards}, which requires the sensor streams to be accessible and linked to processes~\cite{burattin2020mqtt,DBLP:journals/tse/MottolaPOEFFGKM19}.  

The IoT provides new sources of access to real-time data and thus opportunities to develop advanced monitoring approaches, e.g., for condition monitoring, predictive maintenance~\cite{hoppenstedt2018techniques} or verification of process outcomes~\cite{seiger2019toward}. Large amounts of data, different data formats, sampling rates, and data quality are among the challenges that come with using the IoT as a new data source~\cite{de2024interplay,akhramovich2024systematic,kammerer2020process}. With our framework, we focus on using IoT data for event extraction, abstraction, and correlation where we relate raw IoT event streams to the execution of process activities~\cite{10.1007/978-3-030-72693-5_6}. Most works assume rather high levels of process knowledge when discovering activities and processes from sensor data, i.\,e.,~existing activity labels only have to be connected to the raw events~\cite{diba2020extraction,cs6099}. With almost no process knowledge and limited IoT topology knowledge, most approaches are not applicable. 

\subsection{Event Abstraction}

Event abstraction in the context of BPM focuses on the \emph{abstraction gap} between the granularity at which the data is recorded and at which it is analyzed~\cite{janiesch2020internet,diba2020extraction,zerbato2021granularity}. When considering sensor data, the challenge of mapping fine-grained sensor data to more abstract process activities becomes more pronounced~\cite{fi15030113,janiesch2020internet,van2021event}. In literature~\cite{van2021event}, various approaches exist to bridge this abstraction gap using, e.g., Complex Event Processing (CEP)~\cite{seiger2023data,seiger2020towards,soffer2019event} or machine learning (ML)~\cite{folino2014mining,tax2016event,10.1007/978-3-030-72693-5_6}.

\emph{Event abstraction} is needed to map multiple fine-grained sensor data to coarse-grained process events~\cite{van2021event,fi15030113}, e.\,g.,~with the help of CEP~\cite{soffer2019event}, which also supports data pre-processing and context enrichment~\cite{Sztyler2016}; or with machine learning approaches such as clustering~\cite{folino2014mining,fi15030113,ehrendorfer2023clustering} or supervised learning~\cite{tax2016event}. Wanner et al.~combine these two methods based on expert knowledge and observations in the context of a smart factory~\cite{wanner2019countering}.
Then, to be suitable for the application of process mining techniques, such coarse-granular events need to be matched to the process activities in a process model or correlated to the process instances through \emph{event-to-activity mappings}~\cite{BAIER2014123,van2019efficient}. We discuss work about the development of frameworks for IoT processes and event log generation from IoT events that mostly relates to our contribution, and refer to~\cite{soffer2019event,van2021event,de2024interplay,akhramovich2024systematic} for a broader literature overview.

In~\cite{senderovich2016road}, Senderovich et al.~propose a knowledge-driven approach based on interactions mining to transform historical logs of sensor data into standardized event logs including information about process instances.
In~\cite{mannhardt2016low}, Mannhardt et al.~propose a supervised method for event abstraction based on activity patterns. These patterns are derived from low-level events using domain knowledge and then used to build an abstracted event log by exploiting existing alignment techniques. The authors remark the challenges stemming from the distributed, continuous, and ambiguous nature of IoT data, which makes event abstraction a necessary step for event log creation. 
Moreover, Koschmider et al.~\cite{Koschmider2020framework}~discuss a modular framework for obtaining process event logs from sensor data, explicitly considering the data aggregations needed for event correlation, activity discovery, and event abstraction. The lack of explicit process information in the low-level IoT data may lead to \emph{uncertainties} when being correlated to specific cases, as patterns within IoT data can often be associated with multiple activities and processes~\cite{Koschmider2020framework}. Ehrendorfer et al.~present in~\cite{ehrendorfer2019conformance} an approach for conformance checking and classification of manufacturing log data from a smart factory, thereby raising the analysis of low-level IoT data to a process-oriented perspective.

\subsection{Conclusions from Related Work}

From our investigations of related work at the intersection of BPM and IoT, we can conclude that the application of IoT technology in the domain of BPM has received increasing attention over the past years~\cite{janiesch2020internet,de2024interplay}. Several approaches investigate aspects regarding IoT as new data source for enriching process event logs and for supporting process mining~\cite{akhramovich2024systematic}. Our discussions in Section~\ref{sec:awareness} have shown that detecting process-level events and activities from raw IoT data is a multi-step process. Related work either focuses on using IoT data in early stages of activity detection (e.g.,~event extraction), on using specific sensors to detect specific types of activities independent of the process context, or on parts of the event abstraction~\cite{fi15030113}. In later stages of event and activity detection (e.g.,~event abstraction, event correlation, process mining), IoT data plays a less significant role. The goal of this work is to derive a list of challenges that have to be addressed when going the entire way to increase the level of process awareness of IoT data as defined in Section~\ref{sec:awareness} from \emph{no process awareness} to \emph{full process awareness}. From these challenges, which are backed up by additional related work and literature, we intend to develop a holistic framework consisting of a generally applicable sequence of steps that address these challenges in order to increase process awareness of IoT data. Thereby, the framework should serve as a skeleton defining generic, abstract steps and their expected inputs and outputs. The particular techniques and approaches discussed in the aforementioned related work can then be plugged in as concrete mechanisms to implement one or multiple steps of the framework.

\section{Challenges}
\label{sec:challenges}

We have identified the following challenges regarding the automatic connection and derivation of process data and IoT data. These challenges are based on aforementioned related work and additional literature~\cite{janiesch2020internet,Brauner.2022_CSPerspectiveOnDigitalTransformation,Stoiber.2021_EventDrivenBPM,Bertrand.2022_EventLogStandards} as well as on our own experience from implementing real-world use cases regarding the integration of BPM and IoT data in smart factories~\cite{cs6099,10.1007/978-3-030-66498-5_8,SEIGER2022575,9860104,9860216}, smart healthcare~\cite{stertz2020balancing,franceschetti2023proambition}, and smart homes~\cite{seiger2019toward,8432166}.

The most relevant basis for this work is the BPM-IoT manifesto~\cite{janiesch2020internet}, which introduces 16~high-level challenges related to the interaction between the IoT and BPM. The manifesto presents these challenges along a similar spectrum as we have introduced in Section~\ref{sec:awareness}, going from rather low level raw event data to higher level process-related knowledge. We provide a more in-depth view on some specific challenges related to IoT and (business) process data posed in the manifesto. Specifically, we address aspects referring to 1)~\emph{Placing Sensors in a Process-aware Way}; 2)~\emph{Connection of Analytical Processes With the IoT}; 3)~\emph{Integrating the IoT With Process Correctness Checks}; and most prominently 4)~\emph{Bridging the Gap Between Event-Based and Process-Based Systems}~\cite{janiesch2020internet}. With that, we cover all major challenges that are related to collecting and analyzing raw event data from sensors and actuators with the goal of deriving higher-level knowledge. This knowledge is then used for tasks in the BPM context, e.g,~the discovery of new processes, prediction and adaptation of processes, or enacting responses~\cite{janiesch2020internet}. 
\\
\\
\noindent We have identified the following list of 10 challenges:

\setlist[enumerate,1]{start=0}
\begin{enumerate}[label=\textbf{C\arabic*}]

\item \label{itm:c0} Data delivered by IoT sensors is often incompletely described, e.g., the measurement procedure affecting the granularity and the units are missing, and sensors are not aware of their location~\cite{s18010313,hirmer2016automated}. Moreover, missing knowledge about the granularity of sensors hinders the combination and aggregation of different sensors, which is crucial for conducting process mining on IoT data~\cite{Bertrand.2022_EventLogStandards}.

\item \label{itm:c1} Typically, IoT environments produce a plethora of data from heterogeneous sources~\cite{Brauner.2022_CSPerspectiveOnDigitalTransformation}, both regarding volume and variety~\cite{beyel2023analyzing}, which can only be stored and analyzed with considerable computational effort~\cite{kammerer2020process,KUFNER2021103389,Beverungen.2020_SevenParadoxesInBPM}.

\item \label{itm:c2} Data delivered by IoT sensors has different meaning in different scenarios, e.g., a certain temperature reading from one sensor might point towards an error in one scenario/process, while it signifies nominal operation for a second scenario/process~\cite{9860216,hirmer2015sitrs}. Thus, it is desirable that the implemented processes are stronger coupled with the corresponding data produced during execution (cf. Challenge~6 in~\cite{Beverungen.2020_SevenParadoxesInBPM}).

\item \label{itm:c3} Given a scenario and an IoT environment, the complete set of sensors and sensor information required for process analysis cannot be easily derived~\cite{Brauner.2022_CSPerspectiveOnDigitalTransformation,klein2019ftonto}, which is mainly due to the missing integration of process and IoT data within a common event log format or model~\cite{Bertrand.2022_EventLogStandards,https://doi.org/10.48550/arxiv.2206.11392}. This results in potential ambiguities regarding the process--IoT data correlation~\cite{franceschetti2023characterisation,seiger2020towards,mannhardt2016low,Bertrand.2022_EventLogStandards}.

\item \label{itm:c4} Different sensors are relevant for different scenarios, both regarding if they are required at all, and regarding their significance to the scenario/context~\cite{weyers2022,hirmer2015sitrs,Maamar.2020_BridgingTheGapBetweenBPMAndIoT,Valderas.2022_DevelopmentOfIOTEnhancedProcesses}. In addition, specific IoT data can be relevant in more than one scenario (cf.~Challenge 3 -- \emph{Scope of Relevance} in~\cite{Bertrand.2022_EventLogStandards}).

\item \label{itm:c5} Data analysis is often dependent on visualization and human involvement~\cite{weyers2022,weyers2022method}, which makes many of these interactive approaches infeasible, for example for small lot-sizes in manufacturing scenarios or for high-volume, data-intensive production processes~\cite{kammerer2020process}. This leads to data analysis being a highly demanding task and thus process refinements being difficult to achieve (cf. Challenge~9 in~\cite{Beverungen.2020_SevenParadoxesInBPM}).

\item \label{itm:c6} In semi-structured domains with high human involvement, it is challenging to get complete process/event and IoT data in a desirable and uniform granularity~\cite{franceschetti2023proambition,zerbato2021granularity, kammerer2020process,Brauner.2022_CSPerspectiveOnDigitalTransformation}. Moreover, monitoring of human activities quickly raises privacy challenges~\cite{chanal2020security} and is, in general, difficult to achieve \cite{Knoch.2020_VideoToModelHumanActivities}. 
\item \label{itm:c7} Refining existing process scenarios requires in-depth domain knowledge, and is currently neither well-supported by a methodology nor by (semi-)automatic decision support systems~\cite{janiesch2020internet,SEIGER2022575,Valderas.2022_DevelopmentOfIOTEnhancedProcesses}.
The automatic correlation of IoT data streams, process instances, and process models for the purpose of process refinement is currently not possible.

\item \label{itm:c8} In some scenarios, the explicit knowledge of processes might not be complete~\cite{janiesch2020internet,seiger2020towards,Brauner.2022_CSPerspectiveOnDigitalTransformation,Gruhn.2021_BRIBOT}. Processes might be emerging and evolving~\cite{marrella2018supporting,seiger2019toward}. This makes predicting which IoT data is required for analysis difficult.

\item \label{itm:c9} Process outcome prediction, given explicitly known processes, is non-trivial given the high number of influencing IoT devices and data streams as well as physical surroundings (context), thus leading to emergent behavior~\cite{Kopetz2016,janiesch2020internet,Beverungen.2020_SevenParadoxesInBPM}.

\end{enumerate}

With these 10~challenges we provide a comprehensive, though non-exhaustive, more detailed list and discussion--compared to the BPM-IoT manifesto~\cite{janiesch2020internet}--of aspects that emerge when trying to collect and correlate BPM and IoT data to increase process awareness within the IoT. In the following section, we will propose a framework to address these challenges with the goal of moving from no process awareness to full process awareness within a given IoT data set or data stream.

\section{Framework: From No Process Awareness to Full Process Awareness of IoT Data}
\label{sec:contribution}

\begin{figure}
  \centering
  \includegraphics[width=1.0\linewidth]{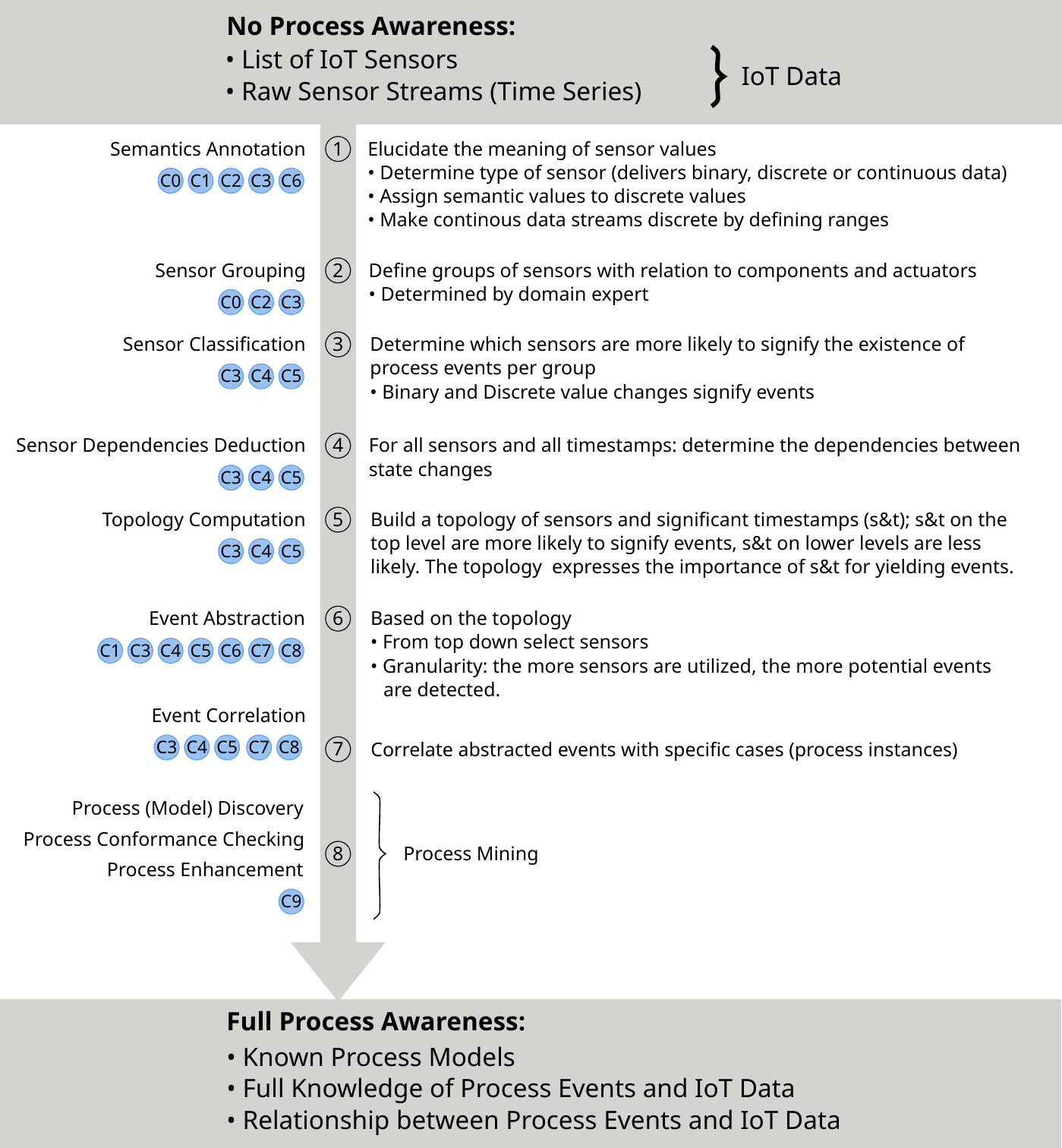}
  \caption{Framework for (semi-)automatic process deduction from IoT data, including references to challenges addressed by each step.}
  \label{fig:layers}
\end{figure}

In this section, we describe a data-driven process deduction framework consisting of eight steps to guide the analysis of a given IoT dataset from no process awareness to full process awareness in accordance with the definitions given in Section~\ref{sec:awareness}, and as depicted in Fig.~\ref{fig:layers}. In the framework we provide a sequence of generally applicable steps and specify their interfaces including their expected inputs and outputs to increase process awareness of the datasets and intermediate artifacts. The goal is to (semi-)automatically transform the raw IoT sensor streams and augment them with process models, knowledge of process events and IoT data, and their relationships. As a starting point, we assume that only the IoT data containing a list of IoT sensors and the respective raw sensor streams are given, i.e., there is no process awareness (cf.~Section~\ref{sec:awareness}). For each step we provide a conceptual description and an instantiation of the framework based on the running example from smart manufacturing (cf.~Section~\ref{sec:example}). We then discuss which of the challenges from Section~\ref{sec:challenges} are addressed by the individual step. Note that the instantiations of the framework's individual steps are intended to show just one possible method or technique to achieve the desired outcome in this step, either relying on the domain expert or already on some (semi-)automated means. Other methods or techniques may be possible to implement the specific steps, as well. The goal of the framework is to streamline the acquisition of process awareness in a series of generalized abstract steps with a high degree of automation potential. The conceptualization and implementation of all possible techniques and algorithms for each step is left for future investigations.

Throughout this section we will continue to use the example introduced in Section~\ref{sec:example}. As already mentioned, how the manufacturing process is
executed / running, is not important: it could be executed through a series of
PLC programs that may or may not include the sensors, through a process engine
enacting an for us unknown process, or through a static Python program that
controls all the machinery.

\subsection{Semantics Annotation}

\subsubsection{Concept:}

In Step~\circled{1} the domain expert elucidates semantics and characteristics
of the sensors such as the type of the sensor values or relations after discretizing continuous streams.
For instance, the process expert can give labels to the values of binary
sensors. These semantics guide the subsequent steps in deducing how changes in
the sensor stream values signify process-level events and thus the execution of
activities. For each sensor the semantics are stored according to the schema
defined in~\cite{https://doi.org/10.48550/arxiv.2206.11392}.

\subsubsection{Challenges:}

Here we address challenges \textbf{\ref{itm:c0}} and \textbf{\ref{itm:c2}}, as
well as \textbf{\ref{itm:c3}} to some extent, by enriching the set of sensor
streams with additional information describing the sensor characteristics to
enable the (semi-)automatic process deduction through subsequent steps. For
\textbf{\ref{itm:c1}}, we assume that through systematically discretizing
continuous data, the amount of stored values can be drastically reduced while
producing equal or better results with less computational effort. The privacy
aspect in \textbf{\ref{itm:c6}} on the other hand can be supported by
annotating data with privacy levels, which can be utilized further on to (a)
always identify when privacy sensitive data is involved, and (b) to regulate
the visibility of such data. This step will be hard to conduct automatically,
as it requires deep semantic knowledge.

\subsubsection{Example:}

Table~\ref{tab:sensorannotation} is the result of the semantic annotations done
by the domain expert for the set of 6 sensors from the smart factory introduced
in the example in Fig.~\ref{fig:running} (cf.~Section~\ref{sec:example}). In our particular example we can see
that all values typically have a semantic meaning. Not being able to map all
continuous sensor values (i.e., ``Source Type'') to discrete values (i.e.,
``Target Type'') is no problem in our framework (as is exemplified in Step 4 --
\emph{Sensor Dependencies Deduction}). But having a clear mapping to discrete values
helps in understanding why certain process events occur, and greatly reduces
the computational complexity of potentially automatically carried out
subsequent steps. From the example it also becomes clear that with a
sufficiently trained AI or Artificial General Intelligence (AGI) this step
could even be conducted automatically.

\begin{table}
\centering
\caption{Sensor annotation, with the source type and target type, and the explanation of the values.}\label{tab:sensorannotation}
\begin{tabular}{|l|l|l|l|l|lllll|}
  \hline
  \multicolumn{5}{|l|}{} & \multicolumn{5}{c|}{\textbf{Value Mapping and Semantics}} \\ \hline
  \textbf{ID} & \textbf{Component} & \textbf{Sensor} & \textbf{Source Type} & \textbf{Target Type} & \textbf{Source} & \textbf{$\rightarrow$} & \textbf{Target} & \textbf{--} & \textbf{Semantics} \\ \hline

  \isquare{1} & Conveyor           & Energy          & continuous           & discrete             & 0               & $\rightarrow$ & 0 & -- & stopped   \\
              &                    & Consumption     &                      &                      & 1..1000         & $\rightarrow$ & 1 & -- & running   \\
              &                    &                 &                      &                      & $>$ 1000        & $\rightarrow$ & 2 & -- & stuck      \\ \hline
  \isquare{2} & Conveyor           & Light Barrier   & binary               & binary               & 0               & $\rightarrow$ & 0 & -- & unbroken  \\
              &                    &                 &                      &                      & 1               & $\rightarrow$ & 1 & -- & triggered \\ \hline
  \isquare{3} & Crane              & Y Movement      & continuous           & discrete             & 0               & $\rightarrow$ & 0 & -- & up        \\
              &                    &                 &                      &                      & 100             & $\rightarrow$ & 1 & -- & down      \\
              &                    &                 &                      &                      & 1..99           & $\rightarrow$ & 2 & -- & moving    \\ \hline
  \isquare{4} & Crane              & X Movement      & continuous           & discrete             & 0               & $\rightarrow$ & 0 & -- & Conveyor pos \\
              &                    &                 &                      &                      & 30              & $\rightarrow$ & 1 & -- & Oven pos     \\
              &                    &                 &                      &                      & 60              & $\rightarrow$ & 2 & -- & Stockpile pos \\
              &                    &                 &                      &                      & 1..29           & $\rightarrow$ & 3 & -- & moving    \\
              &                    &                 &                      &                      & 31..59          & $\rightarrow$ & 3 & -- & moving    \\ \hline
  \isquare{5} & Oven               & Door Status     & binary               & binary               & 0               & $\rightarrow$ & 0 & -- & closed    \\
              &                    &                 &                      &                      & 1               & $\rightarrow$ & 1 & -- & open      \\ \hline
  \isquare{6} & Oven               & Energy          & continuous           & discrete             & 0               & $\rightarrow$ & 0 & -- & off       \\
              &                    & Consumption     &                      &                      & 1..1000         & $\rightarrow$ & 1 & -- & heating   \\
              &                    &                 &                      &                      & $>$ 1000        & $\rightarrow$ & 2 & -- & error     \\ \hline
\end{tabular}
\end{table}

\subsection{Sensor Grouping}

\subsubsection{Concept:}
In Step~\circled{2} the domain expert defines groups of sensors based on the setup of the physical CPS components including their sensors and actuators, which are henceforth also considered as \emph{sensors} producing sensor events (cf.~Fig.~\ref{fig:activitymachinesensor}). This should be rather straight-forward in traditional manufacturing scenarios (e.g.,~grouping for one production machine or one production cell) or medical settings as the grouping is usually done based on physical proximity of sensors and their belonging to one specific CPS component (cf.~Fig.~\ref{fig:activitymachinesensor}). This grouping marks which sets of sensors are impacted by the same activities and should therefore be analyzed together to detect these activities. Groups that only contain one sensor may also exist.

\subsubsection{Challenges:} Here we address the challenges
\textbf{\ref{itm:c0}}, \textbf{\ref{itm:c2}}, and \textbf{\ref{itm:c3}} in a
similar way as in Step~\circled{1}: by further adding information that guides
the subsequent steps. Having a clear semantic relationship-based grouping of sensors
will again help in increasing understandability of how process events could be deducted from sensor
events, as process events are more likely to occur as a
result of interactions \emph{between} sensor groups (i.e., sensor groups are often
equivalent to production cells).

\subsubsection{Example:}

Sensor groups are formed by the domain expert through mostly considering the
location of the sensors and the associated machine. For example, the sensor
group \emph{Buffer} contains different sensors which might encompass one
production cell (cf.~Table~\ref{tab:sensorgrouping}). For the purpose of this
example we added an additional sensor \psquare{7}, which is not part of the
original running example (cf.~Fig.~\ref{fig:running}). \psquare{7} measures
the ambient temperature on the shop floor, and in our case is considered important for the correct functioning of the conveyor and the crane. This exemplifies that \ldots{}

\begin{itemize}

  \item \ldots{} one sensor can be part of multiple groups, and
  \item \ldots{} not all sensors must be part of, or directly associated with a component.

\end{itemize}

\begin{table}
\centering
\caption{Grouping of sensors and assignment of new group identifiers.}\label{tab:sensorgrouping}
\begin{tabular}{|c|c|c|c|c|}
  \hline
  \textbf{Sensor ID} & \textbf{Component} & \textbf{Sensor}    & \textbf{Sensor Group (G)} & \textbf{GID} \\ \hline
  \isquare{1} & Conveyor           & Energy Consumption & Buffer                & B \\ \hline
  \isquare{2} & Conveyor           & Light Barrier      & Buffer                & B \\ \hline
  \isquare{3} & Crane              & Y Movement         & Buffer                & B \\ \hline
  \isquare{4} & Crane              & X Movement         & Buffer                & B \\ \hline
  \psquare{7} & Ambient            & Temperature        & Buffer                & B \\ \hline
  \isquare{5} & Oven               & Door Status        & Oven                  & O \\ \hline
  \isquare{6} & Oven               & Energy Consumption & Oven                  & O \\ \hline
  \isquare{3} & Crane              & Y Movement         & Stockpile             & S \\ \hline
  \isquare{4} & Crane              & X Movement         & Stockpile             & S \\ \hline
  \psquare{7} & Ambient            & Temperature        & Stockpile             & S \\ \hline
\end{tabular}
\end{table}

As can be seen, wherever groups (e.g., production cells) interact, sensors might be part of multiple logical groups (cf.~ambient temperature, crane movement sensors in our example). Defining the correct granularity of the groups will not impact the overall result quality of subsequent analysis steps, but can contribute significantly to visualization and understandability of the results. For our running example we chose the conveyor (representing a buffer), the oven, and the stockpile as the logical groups, but other groups could also make sense. The oven group for example could also profit from containing values from crane sensors.

\subsection{Sensor Classification}

\subsubsection{Concept:}

In Step~\circled{3} each sensor is ranked according to the significance regarding the existence of process-level events (e.g.,~the start or end of an activity) associated with it. Often, sensors with binary value changes are the most significant, followed by sensors with discrete value changes, followed by sensors with continuous value changes~\cite{weyers2022,seiger2020towards}.

While this ordering can be a good first choice, the classification should be confirmed by a domain expert or a more automated technique that analyses sensor data for its relevance. The classification in this step is done per sensor group, whereas inter-group relationships are analyzed in Step~\circled{4}. The resulting classification ranks each sensor per group. A lower ranking sensor (i.e., higher priority) is more likely to indicate its contribution to process-level events. Such a classification can be derived automatically by the information available from steps~\circled{1} and~\circled{2} through analyzing the discretizations, and the respective sensor value stream characteristics within a group.

\subsubsection{Challenges:}

This step addresses challenges \textbf{\ref{itm:c3}, \ref{itm:c4}} and \textbf{\ref{itm:c5}} by semi-automatically deriving a classification signifying the relevance of each sensor per group for the scenario and the data analysis.

\subsubsection{Example:}

The information from the previous steps is analyzed to classify the sensors within a group resulting in Table~\ref{tab:sensorclassdep}.

\begin{table}
\centering
\caption{Ranking of sensors per group (a lower ranking means a higher priority/significance of the sensor).}\label{tab:sensorclassdep}
\begin{tabular}{|c|c|c|c|c|}
  \hline
  \textbf{Sensor ID} & \textbf{GID} & \textbf{Sensor}    & \textbf{Target Type} & \textbf{Sensor Ranking} \\ \hline
  \isquare{1} & B            & Energy Consumption & discrete             & 2 \\
  \isquare{2} &              & Light Barrier      & binary               & 1 \\
  \isquare{3} &              & Y Movement         & discrete             & 2 \\
  \isquare{4} &              & X Movement         & discrete             & 2 \\ \hline
  \isquare{5} & O            & Door Status        & binary               & 1 \\
  \isquare{6} &              & Energy Consumption & discrete             & 2 \\ 
  \isquare{3} &              & Y Movement         & discrete             & 2 \\
  \isquare{4} &              & X Movement         & discrete             & 2 \\ \hline
  \isquare{3} & S            & Y Movement         & discrete             & 1 \\
  \isquare{4} &              & X Movement         & discrete             & 2 \\ \hline
  
\end{tabular}
\end{table}

As can be seen, we propose to rank sensors for each group that was defined in the previous step to simplify the ranking. For example for the first group, the binary sensor has ranking (priority) $1$, while discrete sensors are ranked $2$ (lower priority).

Refined rules might include that the discreteness dimension (binary is a special case of discrete with 2 potential values) is further considered: e.g., X Movement of the crane is considered less important because it has 4 discrete values in comparison to Energy Consumption or Y Movement. Even more refined rules might consider the range of source values, e.g., X Movement is more important because it only spans values from 1 to 59, as compared to the Y Movement which covers values from 0 to 99, or Energy Consumption which covers source values from 0 to greater than 1000.

Finding a good set of rules will be subject of future research, i.e., multiple ranking algorithms with their own strength and weaknesses might emerge.

\subsection{Sensor Dependencies Deduction}

\subsubsection{Concept:}
In Step~\circled{4} dependencies across all sensors and sensor groups are deduced. The goal is to identify important timestamps, where multiple sensors show significant \emph{joint changes}. While discrete sensors might just change their values a few times, continuous sensors typically change their values all the time. In order to identify the importance of a change in sensor values at a particular timestamp, as many as possible sensors, weighted by their importance as calculated in Step~\circled{3}, have to contribute.

This analysis can be done manually by the domain expert, but it also shows high potential for automation based on techniques and metrics for sensor similarity and correlation calculation (e.g.,~using dynamic time warping~\cite{muller2007dynamic}, clustering, or more sophisticated machine learning approaches).

Nevertheless, the concept of detecting joint changes comes
with a set of interesting questions:

\begin{itemize}

    \item In which time frame do joint changes occur? If changes occur within, for example, 1 second, are they joint? How big does the time window need to be?

    \item Should we operate on the original data, or on the discretized data from Step \circled{1}?

    \item If a sensor that provides continuous data (and thus changes all the time) is part of every change, how can we minimize its influence or exclude its dependency to everything?

\end{itemize}

\subsubsection{Challenges:}
This step addresses the same challenges~\textbf{\ref{itm:c3}, \ref{itm:c4}} and \textbf{\ref{itm:c5}} as Step~\circled{3} by revealing dependencies among sensors across groups. The biggest challenge in this step is, that changes do not always happen at the same time, but some sensors might only show a delayed reaction, which makes the change detection for a group of sensors dependent on thresholds and thus computationally expensive.

\begin{figure}
  \centering
  \includegraphics[width=\linewidth]{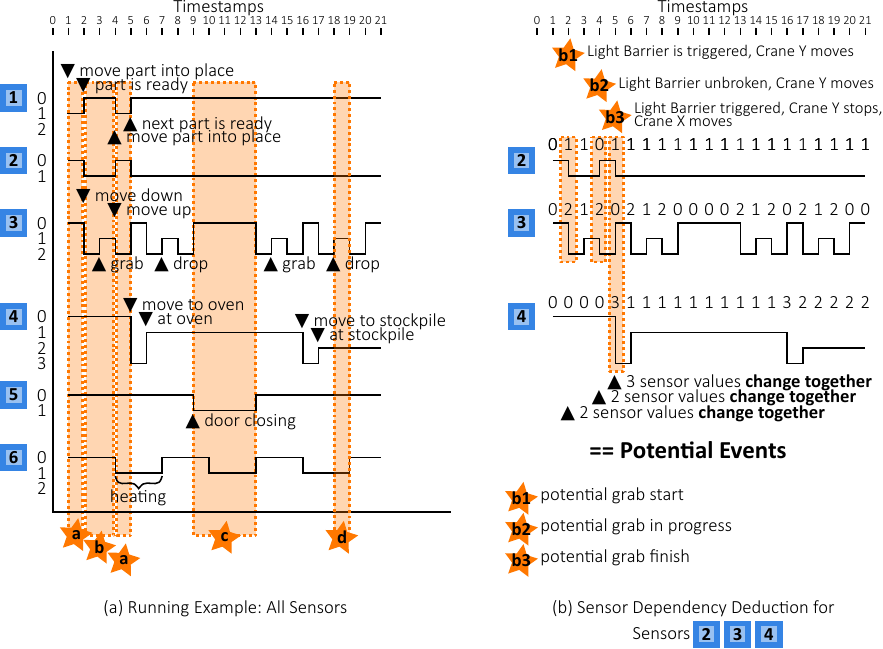}
  \caption{Dependency Deduction: Sensor values change together}
  \label{fig:pattern}
\end{figure}

\subsubsection{Example:}
Sensor dependency deduction infers dependencies between all sensors across groups by means of finding joint changes. Figure~\ref{fig:pattern}(a) shows the discretized changes of all sensors for the complete running example scenario, while Fig.~\ref{fig:pattern}(b) exemplifies the dependencies by
showing how the values change for a set of sensors. Figure~\ref{fig:pattern} also hints at potential solutions for the questions asked above. Joint changes rely on temporal proximity. We presume that by calculating
solutions for multiple time frames (utilizing multiple thresholds), the granularity of dependencies and thus the granularity of potentially detected process events can be controlled. Working with discretized data makes the algorithm much faster, but by pursuing automatic discretization of original data (e.g., by utilizing Symbolic Aggregate approximation~\cite{sun2014improvement} to map to integer values), again the granularity of dependencies and thus the granularity of potentially detected process events can be controlled. Thus a wide range of potential algorithms and optimizations for calculating dependencies seem possible.

For our running example we assume a naive algorithm and rely on the discretized data from Step~\circled{1}. We assume changes happen exactly at the same time. Table \ref{tab:sensordep} shows the changes for all sensors shown in Fig.~\ref{fig:pattern}(b).

\begin{table}
\centering
\caption{Dependencies among sensors.}\label{tab:sensordep}
\begin{tabular}{|c|c|}
  \hline
  { \textbf{Timestamp}} & \multicolumn{1}{l|}{{ \textbf{Joint Change}}} \\
  \hline
  0   & --\\ \hline
  1   & --\\ \hline
  2   & \isquare{2}, \isquare{3}\\ \hline
  3   & \isquare{3}\\ \hline
  4   & \isquare{2}, \isquare{3}\\ \hline
  5   & \isquare{2}, \isquare{3}, \isquare{4}\\ \hline
  6   & \isquare{3}, \isquare{4}\\ \hline
  ... & ... \\ \hline
\end{tabular}
\end{table}

While for Table~\ref{tab:sensordep} we only used sensors \isquare{2}, \isquare{3}, and \isquare{4}, to get the all dependencies, all sensors and all timestamps have to be considered, leading to a large table of timestamps with linked sensor activity.

\subsection{Topology Computation}

\subsubsection{Concept:}
In Step~\circled{5} a list of sensors that represents the sensor \emph{topology}, i.e., the hierarchy and the dependencies among sensors and components and their significance, for a particular process is derived. Sensors are ordered with the goal of having sensors that are more likely to contribute process events on the top of the list, while sensors that are less likely to produce process events (or would produce sensor events at a much finer granularity) at the bottom of the list.

We envision the topology to be automatically derived from information calculated in steps~\circled{3} and~\circled{4}. This topology is intended to
allow process analysts to fine-tune the number of potential process events derived from sensor events, by gradually including sensors from top to bottom of the list.

\subsubsection{Challenges:}
Here we address the same challenges \textbf{\ref{itm:c3}, \ref{itm:c4}} and \textbf{\ref{itm:c5}} as in Steps~\circled{3} and~\circled{4} by computing a topology of sensors in a list. Sensors closer to the top of the list indicate that value changes of these sensors are more relevant to the scenario and the analysis than sensors closer to the bottom.

\subsubsection{Example:}

A naive topology calculation for the event subset \isquare{2}, \isquare{3}, \isquare{4} and the first six timestamps (cf.~Table~\ref{tab:sensordep}, Step \circled{4}) is depicted in Table~\ref{tab:sensortopcalc}.

\begin{table}
\centering
\caption{Sensor topology calculation.}\label{tab:sensortopcalc}
\begin{tabular}{|c|c|c|c|c|c|c|c|c|}
  \hline
  \multicolumn{2}{|l|}{} & \multicolumn{4}{c|}{\textbf{\makecell{Sensor \\  Ranking ($SR_t$)}}} &  \multicolumn{3}{l|}{\textbf{\makecell{Weighted Sensor \\ Ranking ($WSR_t$)}}} \\ \hline
  \textbf{Timestamp $t$} & \textbf{$\alpha_t$} & \isquare{2} & \isquare{3} & \isquare{4} & \textbf{$\Gamma$} & \isquare{2} & \isquare{3} & \isquare{4} \\
  \hline
  0   & $0$ &     &                 &     &     & $0$   & $0$    & $0$ \\ \hline
  1   & $0$ &     &                 &     &     & $0$   & $0$    & $0$ \\ \hline
  2   & $2$ & $1$ & $2$             &     & B   & $2$   & $1$    & $0$ \\ \hline
  3   & $1$ &     & $\frac{2+2}{2}$ &     & B,O & $0$   & $0.5$  & $0$ \\ \hline
  4   & $2$ & $1$ & $2$             &     & B   & $2$   & $1$    & $0$ \\ \hline
  5   & $3$ & $1$ & $2$             & $2$ & B   & $3$   & $1.5$  & $1.5$ \\ \hline
  6   & $2$ &     & $2$             & $2$ & O   & $0$   & $1$    & $1$ \\ \hline \hline
  \multicolumn{5}{|r|}{} & $\Sigma$             & $7$   & $5$  & $2.5$ \\ \hline
  \multicolumn{5}{|r|}{} & $\#$                 & $3$   & $5$    & $3$ \\ \hline \hline
  \multicolumn{5}{|r|}{} & \textbf{$\sfrac{\mathbf{\Sigma}}{\mathbf{\#}}$} & $\mathbf{2.30}$ & $\mathbf{1.00}$ & $\mathbf{0.83}$ \\ \hline
\end{tabular}
\end{table}

For each timestamp $t$ the base importance $\alpha_t$ is set as the number of joint changes, as can be seen in Table~\ref{tab:sensordep}, Step \circled{4}. The sensor ranking $SR_t$ at timestamp $t$ is taken from Table~\ref{tab:sensorclassdep}, Step \circled{3}. As sensors can occur in multiple sensor groups, and can have a different importance in each group, the closest matching group for a combination of sensors is used as the source for the ranking. The sensor group id (GID) is denoted in column $\Gamma$. For example, for timestamp 2, Table~\ref{tab:sensortopcalc} shows $\Gamma = \text{B}$, as sensors \isquare{2} and \isquare{3} only occur together in group B. For timestamp 3, with only sensor \isquare{3} changing, things are more difficult. Sensor \isquare{3} occurs in groups B and O. Thus the weights for both groups have to be averaged (i.e., $\frac{2+1}{2}$). For our naive calculation, the weighted sensor ranking ($WSR_t$) for each timestamp $t$ then amounts to:

\begin{align}
    WSR_t = \alpha_t \times \frac{1}{SR_t}
\end{align}

If a sensor ranking does not exist for a certain timestamp, its value becomes 0. To calculate the importance of a sensor in the topology, the sum ($\Sigma$) of all rankings per sensor is divided by the number of occurrences ($\#$) of a sensor in all timestamps. This results in the final topology (i.e., sorting sensors in a list by importance) as shown in Table~\ref{tab:sensortop}.

\begin{table}

\centering
\caption{Naive Sensor Topology}\label{tab:sensortop}
\begin{tabular}{|c|c|c|}
  \hline
  \textbf{Sensor ID} & \textbf{Importance} & \textbf{Timestamps} \\  \hline
  \isquare{2} & $2.30$               & 2, 4, 5             \\ \hline
  \isquare{4} & $1.00$                 & 5, 6                \\ \hline
  \isquare{3} & $0.83$              & 2, 3, 4, 5, 6       \\ \hline
\end{tabular}
\end{table}

The sensor topology calculation is naive, as similarly to Step \circled{4}, future research into more complex scenarios will determine if alternative calculation algorithms (e.g.,~considering
aspects of the source data) can yield better results.

\subsection{Event Abstraction}

\subsubsection{Concept:}
Steps~\circled{1}--\circled{5} served as event extraction and preparation of event abstraction. In Step~\circled{6} process-level events are abstracted from the information encoded in the topology calculated in Step~\circled{5}~\cite{diba2020extraction}. Due to their higher ranking and thus higher significance to indicate process-level events as explained in the previous steps, we assume that events from sensors at the top of the topology (list) represent process events related to process activities, which group all events from lower levels of the topology.

Just utilizing the top layer (sensors at the top position) of the topology will yield a coarse-grained process model to be mined in Step~\circled{8}, which groups all existing events.
By letting the domain expert specify that the first, second, or even deeper levels, and only a subset of sensors at a particular level should be utilized to determine process activities and thus grouping of events, the process models can be manually tweaked to yield the desired granularity.

\subsubsection{Challenges:}
Here we address the same challenges \textbf{\ref{itm:c3}, \ref{itm:c4}} and \textbf{\ref{itm:c5}} as in the previous three steps by supporting the domain expert in selecting relevant sensors for analysis and event abstraction and automatically extracting the events based on the selection. \textbf{\ref{itm:c1}} can be addressed through having in-depth knowledge about the relationship between different sensors, leading to potential optimization regarding the required data velocity, but also relevance. In other words, data can potentially be discarded because it might not contribute to the data analysis tasks at hand. This step also allows us to control the granularity level of a process, thus for challenge \textbf{\ref{itm:c6}} allowing to stipulate a desired granularity of how human involvement in the process model must be represented. Challenges \textbf{\ref{itm:c7}} and \textbf{\ref{itm:c8}} are directly addressed by the development of this framework,
and realizing Steps \circled{1} to \circled{6}. Generating knowledge and refining knowledge in a structured and semi-automatic way with the goal to refine and evolve processes is the purpose of the proposed framework. For example when the desired granularity of resulting processes is not fine enough, although we utilize the full topology generated in Step \circled{5}, we can deduct that we are missing sensors. On the other hand, the desired granularity of processes can be controlled by the mechanism introduced in Step~\circled{6}.

\subsubsection{Example:}

Beginning at the top of the topology shown in Table~\ref{tab:sensortop}, timestamps 2, 4, and 5 are identified to be points in time when process events occurred.

For our specific sensor subset \isquare{2}, \isquare{3}, \isquare{4} and sensor events for the first six timestamps (cf.~Table~\ref{tab:sensordep}), this yields process events when the light barrier shows changes. These events can be extracted as shown in Table~\ref{tab:eventlog}.

\begin{table}
\centering
\caption{Extracted process events.}\label{tab:eventlog}
\begin{tabular}{|c|l|}%
  \hline
  \bfseries Event timestamp & \bfseries Context \\ \hline
  2 & Group Buffer; Sensors \isquare{2}, \isquare{3} \\ \hline
  4 & Group Buffer; Sensors \isquare{2}, \isquare{3} \\ \hline
  5 & Group Buffer; Sensors \isquare{2}, \isquare{3}, \isquare{4} \\ \hline
\end{tabular}
\end{table}

As can be seen, interpreting the context might require some further insights. When comparing to the large dependency deduction picture shown in Fig.~\ref{fig:pattern}(b), it becomes clear that the three potential events can be interpreted as \emph{pickup} \istar{b} events, but the lifecycle transitions associated with the process events are unclear. Including sensor \isquare{1} could have helped with further differentiation, but it was not included in the sample calculation in Steps \circled{4} and \circled{5}. Nonetheless, the interpretation (i.e., by a human domain expert) shown in Table~\ref{tab:eventloginterpreted} seems feasible.

\begin{table}
\centering
\caption{Abstracted process events.}\label{tab:eventloginterpreted}
\begin{tabular}{|c|c|c|c|}%
  \hline
  \bfseries Process event ID & \bfseries Event timestamp & \bfseries Lifecycle transition & \bfseries Label \\ \hline
  \istar{b1} & 2 & start    & grab part \\ \hline
  \istar{b2} & 4 & unknown  & grab part \\ \hline
  \istar{b3} & 5 & complete & grab part \\ \hline
\end{tabular}
\end{table}

It is important to note that the dependency deduction in Step \circled{4} shows that for timestamp 5 many more sensors are involved.
Thus it is a stronger candidate for a process event. Also the semantic annotation in Step \circled{1} will provide further semantic information about the state of the involved sensors and thus further context for labeling of events and determining the process event lifecycle transition they represent.

Again, future research can show how all the structured information prepared in Steps \circled{1} to \circled{6} can be incorporated in to either (1)~a user interface to support domain experts, or (2)~allow for automatic deduction of process events.

\subsection{Event Correlation}
\subsubsection{Concept:}
In Step~\circled{7} the extracted and abstracted events that were associated with specific process-level activities in Step~\circled{6} are correlated with a specific \emph{case} (i.e., an executed process instance) to enable classic activity-centric process mining approaches~\cite{diba2020extraction}. This step may not be trivial due to the start and end of a case not being clearly identifiable from the IoT data log, and multiple executions of process instances in parallel being possible. Thus, ambiguities may emerge when trying to correlate abstracted events from Step~\circled{6} with a specific case~\cite{franceschetti2023characterisation,li2023rectify}. Additional sensors being annotated to provide data about the start and end of a case, or to track the execution of all activities belonging to one case (e.g.,~a camera~\cite{Malburg.2021_ObjectDetection}) might be necessary to support the correlation. The domain expert has to identify these sensors and specify their correlation with a case, e.g., what is the \emph{object} that the execution of (a) process instance(s) is centered around. This could for instance be the \emph{workpiece} in a production process or the \emph{patient} in a treatment process, which would group all events into an executed case or an object-centric event log~\cite{berti2024ocel}. Moreover, context data from additional sensors and identified preceding process activities could be used to support the event correlation~\cite{brzychczy2024process,fi15030113}.

\subsubsection{Challenges:}

In this step we address challenges \ref{itm:c3}--\ref{itm:c5}, \ref{itm:c7}, and \ref{itm:c8} as process-related data is added here to complement and complete the available information from previous steps. By correlating the extracted and abstracted events with a specific case or object, we achieve the full contextualization of these events within the specific process scenarios.

\subsubsection{Example:}
The domain expert performs the event correlation, i.e., adding the corresponding case identifiers (\emph{Case IDs}) as additional attributes to the activities, by identifying all extracted events as pertaining to the production of the same workpiece and thus belonging to the same process instance (i.e., case). This step is straight-forward under the assumption that no parallel process executions are possible and that all executed production activities relate to the same workpiece (i.e.,~lot size 1)~\cite{weyers2022}. The result of this step is similar to the excerpt of an IoT-enriched event log depicted in Table~\ref{tab:logfull} where all IoT sensor readings are fully contextualized in the specific process and activity instances.

\subsection{Process Mining}

\subsubsection{Concept:}
In Step~\circled{8} process mining techniques can be applied as soon as process activities are identified, which is achieved through \emph{Event Abstraction} with a domain-expert providing a mapping for meaningful labels of the abstracted activities in Step~\circled{6} and the correlation of these activities with specific cases in Step~\circled{7}. Case-based, artifact-based, object-centric, or other process mining techniques can be utilized to mine the processes from the event logs--to discover process models, check conformance, and enhance processes~\cite{van2011process}--and thus to achieve full process awareness of IoT data combined with process-related artifacts (e.g.,~process models).

\subsubsection{Challenges:}

Challenge~\ref{itm:c9} dealing with outcome prediction builds on all the data and relationships generated in Steps \circled{1} to \circled{7}. However, only if the resulting processes are known, potential outcomes can be identified, as they rely on the structure of the process and might only be realized at the very end of a process.

\subsubsection{Example:}

\begin{figure}
  \centering
  \includegraphics[width=1\linewidth]{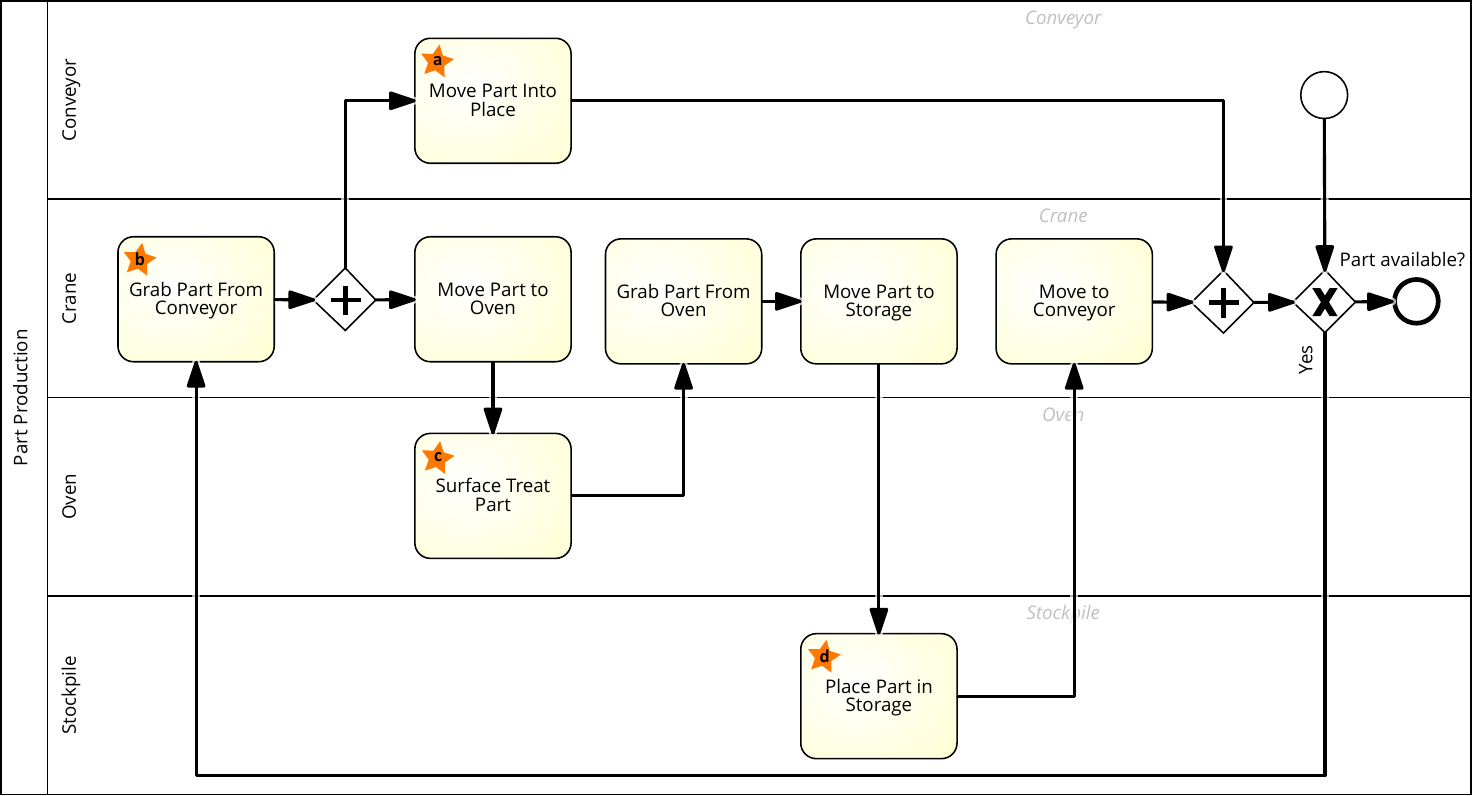}
  \caption{Final process model of the production process in BPMN~2.0.}
  \label{fig:exampleprocesstum}
\end{figure}

Applying the Inductive Miner~\cite{leemans2022robust} on the event log shown excerpt in Table~\ref{tab:eventloginterpreted}--extended with additional entries from all sensors shown in Fig.~\ref{fig:pattern}(a)--but only considering the events with ``start'' lifecycle, the final process model could look as depicted in Fig.~ \ref{fig:exampleprocesstum}. This depends on the process granularity chosen in Step~\circled{6}.

\subsection{Discussion \& Weaknesses}

As implied throughout this chapter, the successful application of this framework relies on a variety of future advances. We will discuss potential problems that have to be solved for each step in the proposed framework. 

If during Step \circled{1} neither domain experts, nor potential AI based annotation mechanisms can provide further insight into the characteristics and meaning of sensor values, it will be very difficult to identify potential process events during Step~\circled{6}. The same holds for Step~\circled{2}, although the physical location of sensors, combined with knowledge about the purpose of sensors might allow for automatic grouping.

Automatic discretization will have to be be explored and can help with reducing human involvement while conducting Step \circled{1} .

We expect Steps \circled{3} to \circled{5} to run fully automatic without further human input, but the algorithms presented as part of this chapter need to be refined. Similar to the multitude of process mining algorithms that exist today, for Steps \circled{3} to \circled{5} we can see multiple refined algorithms dealing with potential shortcomings of topologies.

The abstraction and correlation in Steps~\circled{6} and~\circled{7} will potentially always be conducted by domain experts as they allow fine-tuning the granularity of
the resulting process model(s).

One corner case that has to be mentioned is when the framework is applied to a scenario with only one sensor. Here Step~\circled{1} is the
only step that affects the outcome. If the sensor delivers continuous data, the discretization in Step~\circled{1} will determine all process events, as the topology in Step \circled{5} will contain only this sensor, and all sensor events will be considered to be process events.

Regarding the applicability of the framework in other domains (e.g., healthcare) we do not foresee any problems. As the approach is very basic, depending on just time series data (i.e., timestamps and associated values) from sensors, nothing prevents the application of the approach to data from other domains. For example the approach for activity monitoring from IoT data in healthcare settings described in~\cite{franceschetti2023proambition} exhibits the same characteristics within sensor data relevant for the framework presented in this work, namely time series data and sensor groups.

High volume sensor data will lead to a set of performance problems with the following steps being the most affected:

\begin{itemize}

    \item Step \circled{1} will suffer from high computational cost for discretization, depending on the algorithm for automatic discretization, or the effort needed to transform data.

    \item Step \circled{4} will suffer from performance problems, if a high number of sensors is included in the data, as all sensors have to be compared. The calculation of dependencies for different time windows further multiplies the computational costs.

    \item All other Steps are considered to not be problematic.

\end{itemize}

Furthermore, the development of user interfaces to support domain experts during Steps~\circled{6} and~\circled{7} (and potentially Step~\circled{1}) will be crucial to maximize the utility of the framework, as well as the deduction of process events from IoT events.

\section{Conclusion and Future Work}
\label{sec:conclusion}

Data engineers, data scientists, and domain experts have to be supported in collecting, structuring, and analyzing IoT data to extract process models and event logs in order to (1)~allow potential enhancements to materialize faster, (2)~minimize the errors occurring during the analysis, and (3)~minimize the required domain knowledge. 

In the context of these goals, we identified a set of 10~current research challenges that can be associated with deriving process-relevant data from complex IoT data. To support the goals and address these challenges, we presented a framework consisting of 8~generalized steps that can guide the semi-automated data analysis to produce process data and process models at different levels of granularity starting from raw IoT data (\emph{no process awareness}) up to IoT data which is fully contextualized regarding a process (\emph{full process awareness}). We envision future research to yield a wide variety of results for each step of the framework, including:

\begin{itemize}

    \item Specialized tools for domain experts to annotate and group data artifacts.
    
    \item Custom visualizations to show dependencies between data artifacts in real-time.
    
    \item Refined algorithms to calculate (hidden) dependencies between hardware artifacts and sensor data.
    
    \item Refined data retention policies based on annotations and derived process intelligence.
    
\end{itemize}

To demonstrate the potential of the framework we also introduced a sample scenario and exemplary IoT dataset to discuss how a semi-automatically generated process model based on the steps in the framework can be (a)~of similar quality as a process model provided by domain experts, (b)~used to point out problems in existing process models, and (c)~utilized to refine existing processes, e.g., by pointing out the necessity to introduce additional sensors in a certain context. We think that this framework can serve as means to focus research towards the overall goal of allowing domain experts to make IoT datasets fully process aware without the involvement of data engineers and data scientists~\cite{ehrendorfer2023internet}.

\section*{Acknowledgments}
This research was partly funded by the Austrian Research Promotion Agency (FFG) via the ``Austrian Competence Center for Digital Production'' (CDP) under the contract number 881843. This research was supported by the Pilot Factory Industry 4.0, Seestadtstrasse 27, Vienna, Austria. This work has received funding from the Swiss National Science Foundation under Grant No. IZSTZ0\_208497 (\emph{ProAmbitIon} project).

 \bibliographystyle{splncs04}
\bibliography{bib.bib,intro1.bib,intro2.bib}

\end{document}